\documentclass[a4paper,12pt,english]{article}

\usepackage{helvet}         
\usepackage{courier}        
\usepackage{type1cm}        
\usepackage{makeidx}         
\usepackage{graphicx}        
\usepackage{multicol}        
\usepackage[bottom]{footmisc}

\RequirePackage{hyperref}
\usepackage{cite}
\usepackage{mathtools}
\usepackage{empheq}
\usepackage{url}
\usepackage[mathscr]{euscript}
\usepackage[all,color]{xy}
\usepackage{relsize}
\usepackage{url}
\usepackage{cancel}
\usepackage{caption}
\usepackage[T1]{fontenc}
\usepackage{color}
\usepackage{mathtools}
\usepackage{comment}

\newcommand{\bbm}{\left(\begin{matrix}}
    \newcommand{\ebm}{\end{matrix}\right)}
\newcommand{\beq}{\begin{eqnarray}}
\newcommand{\eeq}{\end{eqnarray}}

\makeatother

\begin{document}

\begingroup
{\let\newpage\relax
\title{Classification of the vacua of the dimensionally reduced low-energy limit of the heterotic string over nearly-K\"ahler manifolds}

\author{G. Manolakos\textsuperscript{1},\,G. Patellis\textsuperscript{2},\,G. Zoupanos\textsuperscript{2,3,4,5}}\date{}
\maketitle}
\begin{center}
\emph{E-mails: giorgismanolakos@gmail.com\,, patellis@central.ntua.gr\,, George.Zoupanos@cern.ch }
\end{center}

\begin{center}
\itshape\textsuperscript{1}Institute of Theoretical Physics, Faculty of Physics, University of Warsaw, ul. Pasteura 5, Warsaw 02-093, Poland\\
\itshape\textsuperscript{2} Physics Department, National Technical
University, Athens, Greece\\
\itshape\textsuperscript{3} Theory Department, CERN\\
\itshape\textsuperscript{4} Max-Planck Institut f\"ur Physik, M\"unchen, Germany\\
\itshape\textsuperscript{5} Institut f\"ur Theoretische Physik der Universit\"at Heidelberg, Germany
\end{center}

%
%
\maketitle

\abstract{We examine the vacua of the scalar potential of the effective 4-d action, obtained after the dimensional reduction of the 10-d $\mathcal{N}=1$ heterotic supergravity coupled to an $\mathcal{N}=1$ Yang-Mills sector. The (Coset Space) dimensional reduction takes place over the three 6-d nearly-K\"ahler manifolds, namely the homogeneous 6-d non-symmetric coset spaces, $G_2/SU(3)$, $Sp_4/SU(2)\times U(1)$ and $SU(3)/U(1)\times U(1)$. The current work consists a complete catalogue of the kinds of vacua of theories obtained after the reduction of the heterotic string over the 6-d non-symmetric coset spaces and, moreover, a contribution to the dialogue of the possibility to result with non-AdS vacua in the framework of string theories.}

\maketitle

\section{Introduction}
With respect to phenomenological viability, the best candidate of all superstring theories \cite{Green:1987sp, Green:1987mn, Lust:1989tj, Polchinski:1998rq, Polchinski:1998rr} is the 10-d, $\mathcal{N}=1$ heterotic string and particularly the version of it that is coupled to the 10-d, $\mathcal{N}=1$ gauge sector of $E_8\times E_8$ \cite{Gross:1985fr}. The latter is favoured compared to the $SO(32)$ alternative version of the heterotic string because of its property that it can be broken down to interesting 4-d Grand Unified Theories (GUTs) which can, in principle, accommodate the Standard Model (SM) gauge group. The aforementioned GUTs are obtained after the employment of appropriate compactification of the extra dimensions. A very strong criterion of the choice of the internal manifold that will serve as the compactification space is the preservation of the initial amount of supersymmetry, in order to achieve contact with the Minimal Supersymmetric Standard Model (MSSM). 

The Calabi-Yau (CY) is an extremely suitable class of compact manifolds that satisfy the requirement of supersymmetry preservation and that is the reason why intensive research has been conducted towards this direction, see e.g. \cite{Witten:1985xb, Derendinger:1985kk} for early attempts. Nevertheless, dimensional reduction over CY manifolds leads to 4-d theories which are accompanied by undesired massless chiral fields (moduli) which, as flat directions of the potential, have values that are not determined (moduli stabilization problem). Besides the moduli stabilization problem, another drawback, not as grave, is that the CY manifolds keep the amount of supersymmetry intact \cite{Candelas:1985en}. At a first glance that is a welcome feature and, as we said above, it was a criterion for the choice of the internal manifolds in the sense that the usual outcome of the dimensional reduction of a higher-dimensional theory is an increase in the amount of supersymmetry. The negative aspect of this preservation of supersymmetry is that, subsequently, it will have to be broken for the sake of phenomenological compatibility and therefore the Soft Supersymmetry Breaking (SSB) sector will have to be introduced externally. 

A very welcome alternative that has been adopted in order that the above difficulties to get tackled, is the employment of a more general class of manifolds, namely those admitting an $SU(3)$-structure, instead of the CY ones. Our interest is focused on a specific type of such manifolds, namely the nearly-K\"ahler manifolds \cite{LopesCardoso:2002vpf, Strominger:1986uh, Lust:1986ix, Castellani:1986rg, Becker:2003yv, Becker:2003sh, Gurrieri:2004dt, Benmachiche:2008ma, Micu:2004tz, Frey:2005zz, Chatzistavrakidis:2008ii, Chatzistavrakidis:2009mh, Dolan:2009nz, Lechtenfeld:2010dr, Klaput:2011mz, Gray:2012md, Klaput:2012vv,Chatzistavrakidis:2008zz}. Specifically in six dimensions, the homogeneous nearly-K\"ahler manifolds are the three non-symmetric coset spaces and the group manifold $SU(2)\times SU(2)$. This type of manifolds have been used in order to attack the moduli stabilisation problem (in the context of flux compactification \cite{Manousselis:2005xa, Popov:2010rf, Chatzistavrakidis:2012qb}), but they also provide another much welcome effect, that is, after the dimensional reduction over them, to result with a 4-d, $\mathcal{N}=1$ effective theory which is not exact but broken, meaning that the SSB terms appear in a natural way, namely the SSB sector is auto-included, well-motivated and its origin is consistently explained in terms of the higher-dimensional theory \cite{Irges:2011de, Manousselis:2001xb, Manousselis:2004xd, Manousselis:2000aj, Manousselis:2001re}.

The dimensional reduction of the low-energy limit of the heterotic string over the 6-d non-symmetric coset spaces mentioned above has already  been examined and has been performed using the Coset Space Dimensional Reduction (CSDR), that is a powerful tool regarding the reduction of higher-dimensional theories in a non-trivial and systematic way \cite{Forgacs:1979zs, Kubyshin:1989vd, Kapetanakis:1992hf}. The explicit implementation of the above reduction of the bosonic part of the heterotic string along with the resulting 4-d effective actions for all three cases of 6-d non-symmetric homogeneous coset spaces can be found in ref.\cite{Chatzistavrakidis:2009mh}.

In the present work, rather than solely focusing on the phenomenologically promising  case of the dimensional reduction of the heterotic string over the coset space $SU(3)/U(1)\times U(1)$, we opt for the complete study of all three above-mentioned non-symmetric coset spaces. For each case, our starting point is the outcome of the reduction, which is a 4-d effective theory, and in this premise we examine the total scalar potential of the theory, which comprises of the gauge, gravity and 3-form sectors. Minimization of the total potential leads to the vacuum of the 4-d theory, as shown explicitly in ref\cite{Chatzistavrakidis:2009mh}. It should be noted that in our previous work \cite{Manolakos:2020cco} we focused on the gauge sector of the effective theory of the $SU(3)/U(1)\times U(1)$ case and our study led to a phenomenologically interesting split-like supersymmetric extension of the SM. As it will be argued, although the present work is independent of any phenomenological model, the restrictions on the parameter space imposed by such a model can have a significant impact on the minimization procedure.   

As expected, the examination of the kinds of vacua that emerge after compactifications of the 10-d $\mathcal{N}=1$ supergravity as a low-energy limit of superstring theories has been a rich field of research for decades, led by experts as well.  First, it was proposed that ``ten to four won't go'', meaning that solutions of the 10-d equations with spontaneous compactification and a maximally symmetric spacetime are ruled out \cite{Freedman:1983zt}. Then, it was proposed that the problem should be addressed in terms that the warped factor is non-trivial, making appropriate a wider class of metrics that are called warped product metrics\footnote{It is a metric that can be almost decomposed into a Cartesian product of manifolds, up to some function depending on the coordinates of the first factor-manifold, multiplying the metric of the second one. In case the warped factor is trivial, the total metric decomposes to a pure product of manifolds.}. Again, it was found that de Sitter vacua were still out of the picture \cite{deWit:1986mwo}. Some years later, a stronger statement came into light with the form of a no-go theorem proving that, there does not exist any non-singular warped compactification of the higher-dimensional theory to Minkowski or de-Sitter space, for $d \geq 2$ with finite $d$ dimensional Newton’s constant, taking into consideration the assumption that the potential of the scalar fields is non-positive \cite{Maldacena:2000mw}. The main desideratum for all the above was the preservation of the $\mathcal{N}=1$ supersymmetry in four dimensions. Later on, a way to avoid the strong no-go theorem was proposed, known as the KKLT mechanism \cite{Kachru:2003aw}. In rough lines, all moduli in the compactification had to be fixed while preserving supersymmetry and after that induce the breaking of supersymmetry in a controlled way by the addition of extra effects (addition of branes in the compactification process), lifting the minimum of the potential in a positive value, resulting with a de Sitter space. 

With the above-mentioned context in mind, let us now explain the state of our case. In our study, we consider the resulting scalar potentials as they result from a Coset Space Dimensional Reduction over the three nearly-K\"ahler manifolds which are not Ricci-flat, but Einstein manifolds, as it is performed on the 10-d $\mathcal{N}=1$ supergravity (low-energy limit of the heterotic string) \cite{Chatzistavrakidis:2009mh}. Therefore, as mentioned above, unlike the CY compactification case, the 4-d resulting theories are not exact supersymmetric; the $\mathcal{N}=1$ supersymmetry is broken by the presence of the additional soft supersymmetry breaking terms involved in the scalar potentials along with the $F-$ and $D-$ terms. Thus, no additional supersymmetry breaking structure is needed and, on top of that, besides the fact that the scalar potential related to the gravity sector is non-positive, the scalar potential expressions related to the gauge and three-form sectors can, in principle, be positive in such a way that may compensate the negative contribution of the gravity sector. Taking all these into account, our study will either result in a viable slice of the parameter space, in which the total potential at the vacuum will be non-negative, or will exclude this possibility and come into terms with the scenaria of AdS vacua.     

The outline of this work is as follows: In Section 2 we recall some basic information on how the low-energy effective action of the $E_8\times E_8$ heterotic string is obtained. We briefly review the procedure of obtaining a 4-d theory via the CSDR procedure in Section 3 and write down the explicit form of the scalar potential of the 4-d theories. In Section 4 we present the main results of our study, namely those of the minimization of the aforementioned scalar potentials. In Section 5, we comment on the case $SU(3)/U(1)\times U(1)$ and, finally, in Section 6 we conclude with highlighting the key points of our analysis.       

\section{Basic preliminaries for the low-energy effective action of heterotic string}
Let us now begin with establishing the context of our work by briefly reminding some relevant key notions from the fundamentals of the string theory which eventually lead to the low-energy theory that is subsequently dimensionally reduced (for complete study see \cite{Green:1987sp, Green:1987mn,Lust:1989tj,Polchinski:1998rq, Polchinski:1998rr,Tong:2009np}). Setting as starting point the action of a string moving in a curved background involving the massless fields $g_{\mu\nu}(X), B_{\mu\nu}(X)$ and $\Phi(X)$, where  $g_{\mu\nu}(X)$ is the metric of the background spacetime, $B_{\mu\nu}(X)$ is the antisymmetric field (Kalb-Ramond), which can be seen as a generalization of the electromagnetic potential (having one more index due to the object propagating is not a point but a string) and $\Phi(X)$ is the dilaton scalar field, it reads:
\begin{align}
    \mathcal{S}&=\frac{1}{4\pi\alpha'}\int \text{d}^2\sigma\left(g_{\mu\nu}\partial_\alpha X^\mu\partial_\beta X^\nu g^{\alpha\beta}+iB_{\mu\nu}\partial_\alpha X^\mu\partial_\beta X^\nu\epsilon^{\alpha\beta}+\alpha'\Phi R\right)\,,
\end{align}
where $\alpha'$ is the Regge slope parameter\footnote{Also it is related to the tension of the string as $\alpha'=1/2\pi T$.}, which is associated to the string length scale as $\alpha'=\ell_s^2$ and obviously $[\alpha']=L^2$, $\sigma^\alpha=(\tau, \sigma)$ are the world-sheet parameters, $g_{\alpha\beta}$ its metric, $R$ the corresponding Ricci scalar and $\epsilon^{\alpha\beta}$ is the antisymmetric 2-tensor. Except for the case in which the dilaton is constant\footnote{This implies that the string coupling constant, $g_s$ is determined by the constant mode of the dilaton, $\Phi_0$, i.e. $g_s=e^{\Phi_0}$. This observation leads to writing the  dilaton field (when non-constant) as a combination of its constant and varying modes, i.e. $\Phi=\Phi_0+\phi$.}, the dilaton term in the worldsheet action violates Weyl invariance and, in order to recover this symmetry after the quantization of the string, one has to observe that the problematic behaviour of the dilaton term can be compensated by a 1-loop contribution of the $g_{\mu\nu}$ and $B_{\mu\nu}$ couplings, since the dilaton term is of first order in $\alpha'$, that is also the expansion parameter of the theory. In order to manifest this compensation, one needs to work with the three $\beta$-functions of the 2-d field theory\footnote{For their detailed expressions see \cite{Tong:2009np}.}, which are obtained by the expressions of the trace of the stress-energy tensor of the theory, and demand that they vanish for the sake of the preservation of Weyl invariance, namely $\beta_{\mu\nu}(G)=\beta_{\mu\nu}(B)=\beta(\Phi)=0$. The crucial argument to obtain the action of the string low-energy limit is to construct it as the action of which the above vanishing of the $\beta$-functions are considered as its corresponding equations of motion for the background space where the string moves. Specifying for the case of the 10-d $E_8\times E_8$ heterotic superstring, the low-energy effective action of the bosonic part will read:
\begin{align}
    \mathcal{S}_b=\frac{1}{2\kappa_0^2}\int \text{d}^{10}x\sqrt{-|g|}e^{-2\Phi}\left(\mathcal{R}-\frac{1}{2}H_{\mu\nu\lambda}H^{\mu\nu\lambda}+4\partial_\mu\Phi\partial^\mu\Phi+ \frac{\alpha'}{4}\text{Tr}\,F_{\mu\nu}F^{\mu\nu}\right)\,,\label{superstringboson}
\end{align}
where $\mu, \nu, \lambda = 0,\ldots 9$, $\kappa_0$ is the coupling constant, $\mathcal{R}$ is the Ricci scalar of the 10-d background space, $F_{\mu\nu}$ is the field strength tensor of $A_\mu$ gauge connection of the $E_8\times E_8$ gauge group, $H_{\mu\nu\lambda}$ is the 3-form field strength tensor of the $B$ field, involving an additional term including the Chern-Simons 3-form built out of the gauge field $A_\mu$, which is related to the absence of anomalies in the heterotic case. The above action is said to be written in the string frame and in order to translate it to (an extension of) the Einstein-Hilbert action it is sufficient to perform a Weyl transformation on the metric:
    $g_{\mu\nu}\longrightarrow g_{\mu\nu}'=e^{\frac{1}{2}\phi}g_{\mu\nu}$.
The above action of the bosonic part, \eqref{superstringboson}, is now written in the Einstein frame as\footnote{Where the constant part of the dilaton field has been absorbed by the coupling constant leading to the redefinition of the latter: $\kappa^2\equiv\kappa_0^2e^{2\Phi_0}=8\pi G_N^{(10)}$, where $[\kappa^2]=L^8$.}:
\begin{equation}
    \mathcal{S}_{b}=\frac{1}{2\kappa^2}\int \text{d}^{10}x\sqrt{-|g|}\left(\mathcal{R}-\frac{1}{2}\partial_\mu\phi\partial^\mu\phi-\frac{1}{12}e^{-\phi}{H}_{\mu\nu\lambda}{H}^{\mu\nu\lambda}+ \frac{\alpha'}{4}e^{-\frac{1}{2}\phi}\text{Tr}\,F_{\mu\nu}F^{\mu\nu}\right)\,.\label{lowenergyheterotic}
\end{equation}
Besides the above action which is the bosonic part of the 10-d $E_8\times E_8$ heterotic string, the complete theory also includes a fermionic and an interacting sector \cite{Bergshoeff:1981um}.

\section{Dimensional Reduction over the three cosets and the scalar potential of the theory}

In order to result with a 4-d theory and make touch with low-energy physics (scales of experiment), the above action has to be dimensionally reduced. As argued in the introduction, we focus on the three cases in which the dimensional reduction is performed over the three 6-d coset spaces.
According to the systematic analysis followed in \cite{Chatzistavrakidis:2009mh}, the Coset Space Dimensional Reduction procedure leads to a specific expression of the scalar potential in four dimensions (the fermionic part is irrelevant in this context). For each of the three cases, the contributions of the three sectors (gravity, 3-form and gauge) of the bosonic part of the heterotic string, eq.\eqref{lowenergyheterotic}, have been examined separately. It is of high importance to recall that the CSDR procedure breaks the initial $G = E_8$ gauge group to a subgroup, $H$, following the rule $H = C_G(R)$, where $R$ is the isotropy group of the coset space over which the reduction is performed. In other words, the specific reduction procedure leads to a broken 4-d gauge symmetry, parametrized by the subgroup $H$ of $G$, which is determined by the centralizer of the isotropy group $R$ of the coset into the initial gauge group $G$ \cite{Kapetanakis:1992hf, Chatzistavrakidis:2009mh}. For the three cases, applying the rule $H = C_G(R)$, the 4-d subgroup will be:
\begin{itemize}
    \item $S/R = G_2/SU(3)$: \[E_8\supset SU(3)\times E_6\,\Rightarrow H = E_6\]
    \item $S/R = Sp_4/SU(2)\times U(1)$: \[E_8\supset SU(3)\times E_6\supset SU(2)\times U(1)\times E_6\,\Rightarrow H = E_6\times U(1)\]
    \item $S/R = SU(3)/U(1)\times U(1)$: \[E_8\supset SU(3)\times E_6\supset SU(2)\times U(1)\times E_6\supset U(1)\times U(1)\times E_6\,\Rightarrow H = E_6\times U(1)\times U(1)\]
\end{itemize}

Here, we list the expressions of the scalar potential of the three sectors in all three cases in four dimensions as they have been obtained in ref \cite{Chatzistavrakidis:2009mh}. 

\subsection{The $G_2/SU(3)$ case}
The expressions of the three sectors composing the total scalar potential are given as\footnote{It is worth-noting that the whole expression of the gravitational component of the potential is calculated in the Einstein frame, that is why the dilaton scalar field is present.}:
\begin{align}
    V_{grav} &= -\frac{15}{\kappa^2}\frac{e^{-\tilde{\phi}}}{R_1^2}\label{gravitysectorcasei}
\end{align}
\begin{align}
    V_{3f} &= \frac{1}{\kappa^2}e^{-\tilde{\phi}}\left[ \frac{b^2}{R_1^6} + \frac{\sqrt{2}}{R_1^3}i\alpha^\prime b(d_{ijk}\beta^i\beta^j\beta^k - h.c.) + 2\alpha^{\prime 2}\beta^i\beta^j\beta^kd_{ijk}d^{lmn}\beta_l\beta_m\beta_n \right.\nonumber \\
    &+ \left. \frac{3}{R_1^2}\alpha^{\prime 2}\beta^4 - \frac{\sqrt{6}}{R_1}\alpha^{\prime 2}\beta^2(d_{ijk}\beta^i\beta^j\beta^k + h.c.)\right]\label{3formsectorcasei}
\end{align}
\begin{align}
    V_{gauge} &= \frac{\alpha^\prime}{8\kappa^2}e^{\frac{\tilde{\phi}}{2}}\left( \frac{8}{R_1^2}-\frac{40}{3R_1^2}\beta^2 - \left[\frac{4}{R_1}d_{ijk}\beta^i\beta^j\beta^k + h.c.\right]\right.\nonumber\\
    & +\left. \beta^i\beta^jd_{ijk}d^{klm}\beta_l\beta_m + \frac{11}{4}\sum_{\alpha}\beta^i(G^\alpha)_i^j\beta_j\beta^k(G^\alpha)_k^l\beta_l\right)\,,\label{gaugesectorcasei}
\end{align}
where $H=E_6$, $d_{ijk}$ is the symmetric $E_6$ invariant tensor, $G^\alpha$ are the $78$ generators of the $E_6$ group in the fundamental representation, $\beta^i$ are the scalar components of chiral superfields (27plets), $\beta^2 = \beta_i\beta^i$ and $b$ is a parameter related to the kinetic term, $H$, of the 2-form field, $B$ \footnote{In a few words, the $b^i$s appearing in the three cases are the parameters accompanying the expansion forms $\omega^i(y)$ which are the S-invariant 2-forms on the internal space (for more details see \cite{Chatzistavrakidis:2009mh}).}. Also, $R_1$ is the radius of the coset:
\begin{equation}
    R_1(\phi, \varphi) = \frac{r}{\sqrt{3}}e^{-\frac{\tilde{\varphi}(\phi, \varphi)}{2\sqrt{3}}}\,,\label{radiuscasei}
\end{equation}
where $r$ is a constant value around which the radius is fluctuating due to the presence of the modulo field and, as written down explicitly, it is a function of the scalar fields, $\phi, \varphi$, which are implicitly introduced through the following redefinitions:
\begin{align}
    \tilde{\phi} &= \tilde{\phi}(\phi, \varphi) = \frac{1}{2}(-\phi + \sqrt{3}\varphi)\,,\nonumber\\
    \tilde{\varphi} &= \tilde{\varphi}(\phi, \varphi) = \frac{1}{2}(-\varphi - \sqrt{3}\phi)\,.\label{scalarscasei}
\end{align}

\subsection{The $Sp_4/SU(2)\times U(1)$ case}
The expressions of the three sectors composing the total scalar potential are given as:
\begin{align}
    V_{grav} &= -\frac{1}{4\kappa^2}e^{-\tilde{\phi}}\left(\frac{4}{R_2^2} + \frac{12}{R_1^2} - \frac{R_2^2}{R_1^4}\right)\label{gravitysectorcaseii}
\end{align}
\begin{align}
    V_{3f} &= \frac{1}{4\kappa^2}e^{-\tilde{\phi}}\left[\frac{1}{(R_1^2R_2)^2}(2b^1 + b^2)^2 + \sqrt{2}i\alpha^\prime \frac{1}{R_1^2R_2}(2b^1+b^2)(d_{ijk}\alpha^i\alpha^j\beta^k - h.c.)\right. \nonumber\\
    &+ \left. 8\alpha^{\prime 2}\alpha^i\alpha^j\beta^k d_{ijk}d^{lmn}\alpha_l\alpha_m\beta_n + \alpha^{\prime 2} \left(\frac{\alpha^2}{R_1} + \frac{\beta^2}{R_2}\right)^2\right. \nonumber\\
    &\left. +\sqrt{6}\alpha^{\prime 2}\left(\frac{\alpha^2}{R_1} + \frac{\beta^2}{R_2}\right)(d_{ijk}\alpha^i\alpha^j\beta^k + h.c.)\right]\label{3formsectorcaseii}
\end{align}
\begin{align}
    V_{gauge} &= \frac{\alpha^\prime}{8\kappa^2}e^{-\frac{\tilde{\phi}}{2}}\left[12\left(\frac{1}{R_1^4} + \frac{1}{R_2^4}\right) - \frac{6}{R_1^2}\alpha^2 - \frac{4}{R_2^2}\beta^2 \right . \nonumber \\
    &+ \left(4\sqrt{\frac{10}{7}}R_2\left(\frac{1}{R_2^2} + \frac{1}{2R_1^2}\right)d_{ijk}\alpha^i\alpha^j\beta^k+h.c.\right)
    \nonumber\\
    &+\left.6\left(\alpha^i(G^\alpha)_i^j\alpha_j + \beta^i(G^\alpha)_i^j\beta_j\right)^2 + \frac{1}{3}\left(\alpha^i\alpha_i - 2\beta^i\beta_i\right)^2
    \right.\nonumber\\
    &+\left. \frac{5}{7}\alpha^i\alpha^jd_{ijk}d^{klm}\alpha_l\alpha_m + \frac{20}{7}\alpha^i\beta^jd_{ijk}d^{klm}\alpha_l\beta_m\right]\,,\label{gaugesectorcaseii}
\end{align}
where every symbol is the same as in the previous case and, furthermore since $H=E_6\times U(1)$, $\alpha^i$ is another $E_6$ chiral 27plet and $\alpha, \beta$ are $E_6$ chiral singlets bearing only $U(1)$ charge. The radii are given as: 
\begin{align}
    R_1^2(\phi, \varphi, \chi) &= r^2e^{-\frac{\tilde{\varphi}(\phi, \varphi, \chi)}{\sqrt{2}}}\nonumber\\
    R_2^2(\phi, \varphi, \chi) &= r^2e^{-\tilde{\chi}(\phi, \varphi, \chi)}\,,\label{radiuscaseii}
\end{align}
and the scalar fields are combined to give the following redefinitions:
\begin{align}
    \tilde{\phi} &= \tilde{\phi}(\phi, \varphi) = -\frac{1}{2}(\phi - \sqrt{3}\varphi)\,,\nonumber\\
    \tilde{\varphi} &= \tilde{\varphi}(\phi, \varphi, \chi) = -\frac{\sqrt{2}}{2}(\phi + \frac{1}{\sqrt{3}}\varphi + 4\gamma\chi)\,,\nonumber\\
    \tilde{\chi} &= \tilde{\chi}(\phi, \varphi, \chi) = -\frac{1}{2}(\phi + \frac{1}{\sqrt{3}}\varphi - 8\gamma\chi)\,,\label{scalarscaseii}
\end{align}
where $\gamma^2 = 1/24$.

\subsection{The $SU(3)/U(1)\times U(1)$ case}

The expressions of the three sectors composing the total scalar potential are given as:
\begin{align}
    V_{grav} &= -\frac{1}{4\kappa^2}e^{-\tilde{\phi}}\left(\frac{6}{R_1^2}+\frac{6}{R_2^2}+\frac{6}{R_3^2}-\frac{R_1^2}{R_2^2R_3^2}-\frac{R_2^2}{R_1^2R_3^2}-\frac{R_3^2}{R_1^2R_2^2}\right)\label{gravitysectorcaseiii}
\end{align}
\begin{align}
    V_{3f} &= \frac{1}{4\kappa^2}e^{-\tilde{\phi}}\left[\frac{(b_1^2+b_2^2+b_3^2)^2}{(R_1R_2R_3)^2}+\sqrt{2}i\alpha'\frac{1}{R_1R_2R_3}(b_1^2+b_2^2+b_3^2)(d_{ijk}\alpha^i\beta^j\gamma^k-h.c.\right.\nonumber\\
    &\left.+ 8\alpha^{\prime 2}\alpha^i, \beta^j, \gamma^kd_{ijk}d^{lmn}\alpha_l\beta_m\gamma_n + \alpha^{\prime 2}\left(\frac{\alpha^2}{R_1} + \frac{\beta^2}{R_2} + \frac{\gamma^2}{R_3}\right)^2\right.\nonumber\\
    &\left.+\sqrt{6}\alpha^{\prime 2} \left(\frac{\alpha^2}{R_1} + \frac{\beta^2}{R_2} + \frac{\gamma^2}{R_3}\right)(d_{ijk}\alpha^i\beta^j\gamma^j + h.c.) \right]\label{3formsectorcaseiii}
\end{align}
\begin{align}
    V_{gauge} &= \frac{\alpha'}{8\kappa^2}e^{-\frac{\tilde{\phi}}{2}}\left[ \frac{2}{5}\left(\frac{1}{R_1^4} + \frac{1}{R_2^4} + \frac{1}{R_3^4} \right) + \left(\frac{4R_1^2}{R_2^2R_3^2}-\frac{8}{R_1^2}\right)\alpha^i\alpha_i \right. \nonumber\\
    &+\left(\frac{4R_2^2}{R_1^2R_3^2}-\frac{8}{R_2^2}\right)\beta^i\beta_i+\left(\frac{4R_3^2}{R_1^2R_2^2}-\frac{8}{R_3^2}\right)\gamma^i\gamma_i   \nonumber\\
    &+\sqrt{2}80\frac{R_1^2+R^2_2+R_3^2}{R_1R_2R_3}(d_{ijk}\alpha^i\beta^j\gamma^k+h.c.)\nonumber\\
    &+\frac{1}{6}\left(\alpha^i(G^\alpha)_i^j\alpha_j+\beta^i(G^\alpha)_i^j\beta_j+\gamma^i(G^\alpha)_i^j\gamma_j\right)^2\nonumber\\
    &+5\left(\alpha^i\alpha_i-\beta^i\beta_i\right)^2+\frac{10}{3}\left(\alpha^i\alpha_i+\beta^i\beta_i-2\gamma^i\gamma_i\right)^2\nonumber\\
    &\left.+40\alpha^i\beta^jd_{ijk}d^{klm}\alpha_l\beta_m+40\beta^i\gamma^jd_{ijk}d^{klm}\beta_l\gamma_m+40\alpha^i\gamma^jd_{ijk}d^{klm}\alpha_l\gamma_m\right]\,,\label{gaugesectorcaseiii}
\end{align}
where the various symbols denote the same quantities as in the previous cases and, furthermore since $H=E_6\times U(1)\times U(1)$, $\gamma^i$ and $\gamma$ are the third chiral 27plet and singlet fields, respectively. Also, $R_1, R_2, R_3$ are the three radii of the internal space where their corresponding expressions with respect to the scalar fields are:
\begin{align}
    R_1^2(\phi, \varphi, \chi, \psi) &= r^2e^{-\tilde{\varphi}(\phi, \varphi, \chi, \psi)}\nonumber\\
    R_2^2(\phi, \varphi, \chi, \psi) &= r^2e^{-\tilde{\chi}(\phi, \varphi, \chi, \psi)}\nonumber \\
    R_3^2(\phi, \varphi, \chi, \psi) &= r^2e^{-\tilde{\psi}(\phi, \varphi, \chi, \psi)} \,.\label{radiuscaseiii}
\end{align}
In this case there are four scalar fields, namely the dilaton $\phi$, and the threee radius moduli, $\varphi, \chi, \psi$. Nevertheless, their more proper expressions that are used throughout the calculations are given by the following combinations:
\begin{align}
    \tilde{\phi} &= \tilde{\phi}(\phi, \varphi) = -\frac{1}{2}(\phi - \sqrt{3}\varphi)\,,\nonumber\\
    \tilde{\varphi} &= \tilde{\varphi}(\phi, \varphi, \chi, \psi) = -\frac{1}{2}(\phi + \frac{1}{\sqrt{3}}\varphi + 4\gamma\chi + 4\delta\psi)\,,\nonumber\\
    \tilde{\chi} &= \tilde{\chi}(\phi, \varphi, \chi, \psi) = -\frac{1}{2}(\phi + \frac{1}{\sqrt{3}}\varphi + 4\gamma\chi - 4\delta\psi)\,,\nonumber\\
    \tilde{\psi} &= \tilde{\psi(\phi, \varphi, \chi, \psi}) = -\frac{1}{2}(\phi+\frac{1}{\sqrt{3}}\varphi -8\gamma\chi)\label{scalarcaseiii}
\end{align}
where $\gamma^2 = 1/24$ and $\delta^2 = 1/8$.

\section{Minimization of the potential - The vacuum}

In each case we will provide the necessary information regarding the computation of the minimization. First, we give a convenient redefinition of the various scalar fields (dilaton and scalars related to the radii), then we write down the expression for the radius in the special nearly-K\"ahler limit in which all radii are equal (the general one is given in \cite{Chatzistavrakidis:2009mh}), then we write down the three components of the total scalar potential (gravity, 3-form and gauge), then the expression of the total potential and finally we give the results related to the minimization of the latter. 

Before we begin this case study, a very crucial consideration we took is related to the gauge potential in all three cases originating from the study of the CSDR procedure. In all cases the initial gauge group $G = E_8$ breaks to a subgroup $H$ (as pointed out in the beginning of the previous section) and in addition we have chosen for simplicity that $S\subset G$ as $E_8\supset S\times K$ in the present examination. Then the subgroup $H$ of $E_8$, in turn, breaks spontaneously to $K$ \cite{Harnad:1978vj, Harnad:1980fz, Kapetanakis:1992hf}. This spontaneous breaking occurs in all three cases and the direction acquiring the vacuum expectation value (vev) suggests the surviving components and facilitates our calculations. We will come back to this in each case separately in the study that follows.

Also, regarding the various constants that are encountered in the calculations we take into account the following:
First, regarding the gauge sector, in all three $V_{gauge}$ expressions we compare to the well-known Yang-Mills analogue with coupling constant $g$ as in \cite{Polchinski:1998rr}:
\begin{align}
    \frac{\alpha^\prime}{8\kappa^2} = \frac{1}{2g^2}\,\Rightarrow\, g^2 = \frac{4\kappa^2}{\alpha^\prime}\,.\label{comparetoyangmills}
\end{align}
Moreover, since:
\begin{equation}
    \kappa^2 = 8\pi G_N\,,
\end{equation}
combined with the above, one results with \cite{Mohapatra:1986uf}:
\begin{equation}
    G_N = \frac{1}{8}a_u\alpha^\prime\,.
\end{equation}
Now, using the generic unification gauge coupling, $a_u = 1/24$ and since Newton's constant in four dimensions is known, one can end up with the Regge slope, $\alpha^\prime \simeq 1.29 \cdot 10^{-36} GeV^{-2}$ and consequently with the string scale, $M_{str} = 8.8 \cdot 10^{17} GeV$. In turn, in order to obtain the compactification scale, we start with the 10-d gravitational coupling $\kappa_{10}^2 = \kappa_0^2e^{2\Phi_0}\approx M_{str}^{-8}e^{2\Phi_0}$, which is connected to the 4-d by $\kappa^2 = \kappa_{10}^2/V$, where $V$ is the volume of the coset space, let us pick, for instance, the $SU(3)/U(1)\times U(1)$ which has $V = \frac{\pi^3}{2}R_1^2R_2^2R_3^2=\frac{\pi^3}{2}M_C^{-6}$. Calculations lead to the following result:
\begin{equation}
    M_C = \sqrt[6]{\frac{4\pi^4 M_{str}^8}{e^{2\Phi_0}M_pl^2}}\sim M_{str}\,.
\end{equation}
Also, since we consider $e^{2\Phi_0} = g_s = g_u = 0.7236$, we end up with $\Phi_0\sim -0.2$. Therefore, in the ensuing, we take the varying part of the dilaton to be negative and of the order of $\mathcal{O}(10^{-1})$. We can take it even smaller without any qualitative change of our results.

\subsection{The $G_2/SU(3)$ case}
This case is genuinely nearly-K\"ahler, which means that there is no need to take the corresponding limit. The two scalar fields, i.e. the dilaton $\phi$ and the radius scalar $\varphi$, are found in eq.\eqref{scalarscasei} and the radius of the space, $R_1$, is given in eq.\eqref{radiuscasei}. \\\\

\newpage

\noindent\underline{\emph{The gravitational sector:}} \\\\
From eq.\eqref{gravitysectorcasei}, the (negative-definite) contribution of the potential due to the gravitational sector is:
\begin{equation}
    V_{gr}(\phi, \varphi) = -\frac{15}{\kappa^2}\frac{e^{-\tilde{\phi}(\phi,\varphi)}}{R_1(\phi,\varphi)^2}\,,
\end{equation}
where $\kappa$ is the gravitational coupling in four dimensions, related to the 10-d one by $\kappa^2 \equiv \kappa_4^2 = \kappa_{10}^2/vol_6$. 
\\\\
\underline{\emph{The gauge sector:}} \\\\
Considering $S/R = G_2/SU(3)$, the $R=SU(3)$ is embedded into $G = E_8$ as:
\begin{align}
E_8&\supset SU(3)\times E_6\\
248 &= (8,1) + (1,78) + (3,27) + (\bar{3}, \bar{27})
\end{align}
and according to the rule $G\supset R\times H$, the initial gauge group $E_8$ breaks to the subgroup $H=E_6$. In turn, since we have chosen that $S\subset G$ as:
\begin{align}
E_8&\supset G_2\times F_4 \\
248 &= (14,1) + (1,52) + (7,26)
\end{align}
the subgroup $H=E_6$, in turn, breaks spontaneously to the $K=F_4$ under the following decomposition:
\begin{equation} 
E_6\supset F_4\,,\quad 27=1+26\,.
\end{equation}
Therefore the value of the 27plet that breaks the symmetry of $H=E_6$ to $K=F_4$ is given at the singlet direction (that is the 1st one, $i=1$), does not break further the $F_4$ subgroup. Therefore, in the following calculations, we keep only the singlet direction out of the entire 27plet of the $E_6$ and, starting from eq.\eqref{gaugesectorcasei}, the contribution to the potential due to the gauge sector is:
\begin{equation}
    V_{gauge}(\phi,\varphi,\beta) = \frac{\alpha^{\prime}}{8\kappa^2}e^{-\frac{\tilde{\phi}(\phi,\varphi)}{2}}\left(\frac{8}{R_1(\phi,\varphi)^4} - \frac{40\beta^2}{3R_1(\phi,\varphi)^2} + \frac{11}{6}\beta^4 \right)\,,
\end{equation}
where $\beta$ is the scalar component of the 27plet chiral  superfield that acquires vev after the spontaneous breaking of the subgroup $H=E_6$ to $K=F_4$.
\\\\
\underline{\emph{The 3-form sector:}} \\\\
From, eq.\eqref{3formsectorcasei}, the contribution to the potential due to the 3-form part is:
\begin{equation}
    V_{3f}(\phi,\varphi,\beta) = \frac{1}{\kappa^2}e^{-\tilde{\phi}(\phi,\varphi)}\left(\frac{b^2}{R_1(\phi,\varphi)^6} + \frac{3\alpha^{\prime 2}\beta^4}{R_1(\phi,\varphi)^2}\right)\,.
\end{equation}
Putting the above three contributions together, the total scalar potential is found:
\begin{align}
    V &= \frac{e^{-\sqrt{3} \varphi -\phi}}{48 \kappa ^2 r^6} \left(1296 b^2+r^2 e^{\frac{\varphi }{\sqrt{3}}+\frac{\phi }{4}}\right.\nonumber\\ 
    &\left.\left(432 \alpha^\prime  e^{\frac{\varphi }{4 \sqrt{3}}}+11 \alpha^\prime  \beta ^4 r^4 e^{\frac{5 \varphi }{4 \sqrt{3}}+\phi }+432 r^2 e^{\frac{3 \phi }{4}} \left(\alpha ^{\prime 2} \beta ^4-5\right)-240 \alpha^\prime  \beta ^2 r^2 e^{\frac{1}{4} \left(\sqrt{3} \varphi +2 \phi \right)}\right)\right)\,.
\end{align}
Minimization of the above scalar potential with respect to the two fields, the scalar field $\varphi$  related to the radius and the vev-acquiring scalar component of the chiral superfield $\beta$, gives the the scalar potential at the vacuum for the various values of the $b$ parameter.\\

\begin{minipage}[b]{0.4\linewidth}\centering
\includegraphics[width=0.95\textwidth]{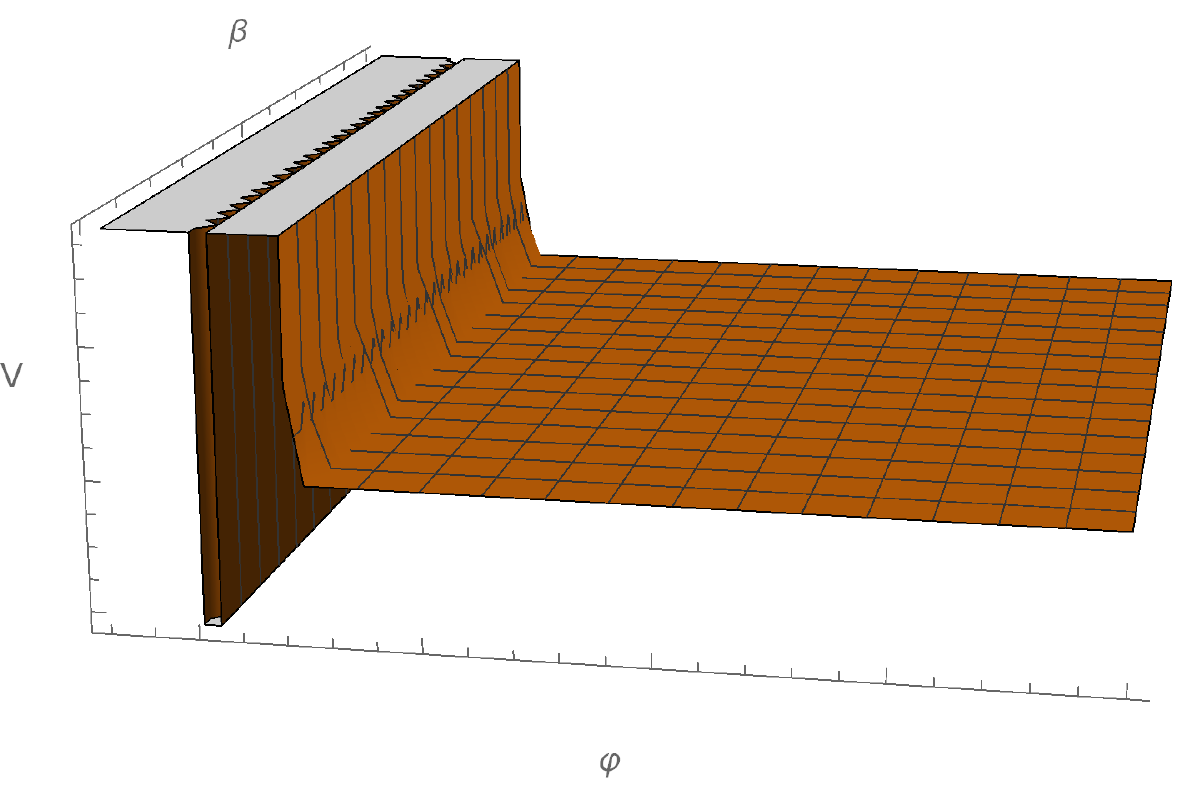}\\
\includegraphics[width=\textwidth]{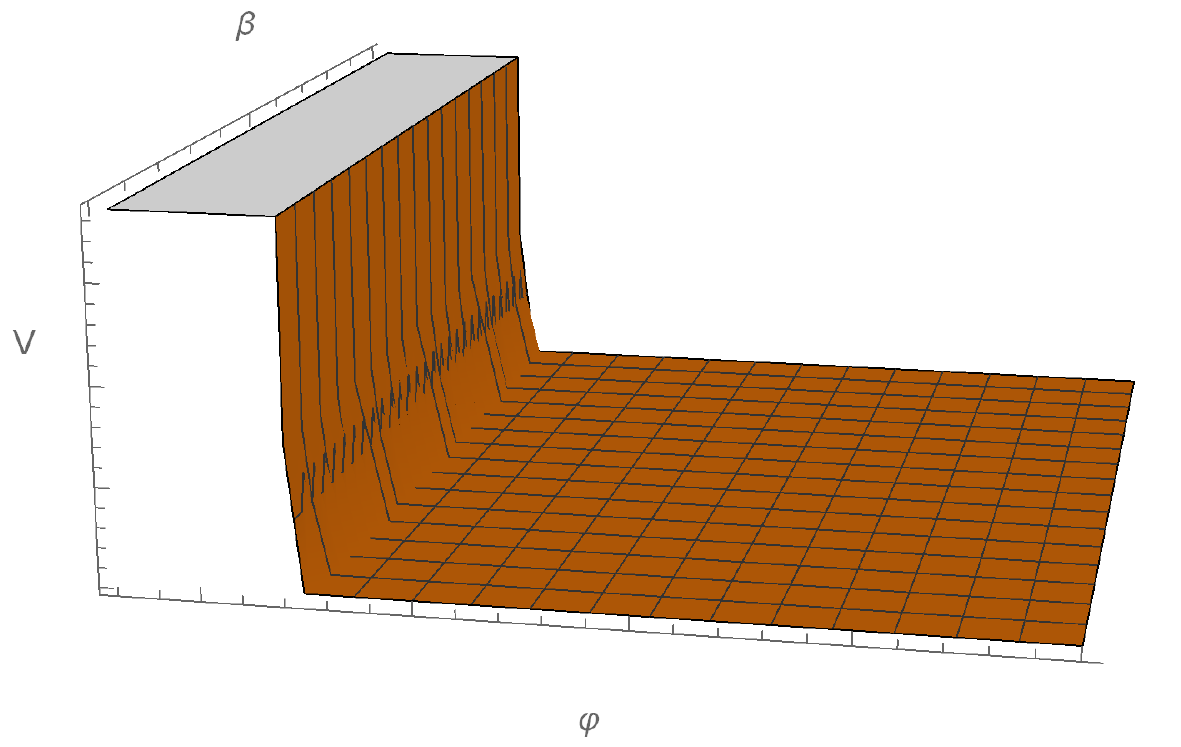}\captionof{figure}{\smaller Top: Qualitative presentation of the potential  for $b<10^{-35}$ for the $G_2/SU(3)$ case.\\
Bottom: The potential for the same case for $b>10^{-35}$.}\label{fig:case1}
\end{minipage}\hspace{0.05\textwidth}
\begin{minipage}[b]{0.4\textwidth}\centering
\begin{tabular}{|l|r|r|r|}
\hline
$b$ & $\varphi$ & $<\beta>$ & $V$ \\\hline
$10^{-42}$ & $-55.7$ & $-85.3$ & $-7.3 \cdot 10^{101}$  \\\hline
$10^{-41}$ & $-47.7$ & $-74.0$ &  $-7.3 \cdot 10^{97}$ \\\hline
$10^{-40}$ & $-31.7$ & $-49.4$ &  $-7.3 \cdot 10^{89}$ \\\hline
$10^{-39}$ & $-23.8$ & $-36.1$ &  $-7.2 \cdot 10^{85}$ \\\hline
$10^{-38}$ & $-15.8$ & $-24.5$ &  $-6.8 \cdot 10^{81}$ \\\hline
$10^{-37}$ & $-7.8$ & $-12.0$ &  $-5.8 \cdot 10^{77}$ \\\hline
$10^{-36}$ & $0.6$ & $1.1$ &  $-3.0 \cdot 10^{73}$ \\\hline
$10^{-35}$ & $7482.7$ & $13867.7$ &  $0$ \\\hline
$10^{-34}$ & $7482.7$ & $13867.7$ &  $0$ \\\hline
$10^{-33}$ & $7482.7$ & $13867.7$ &  $0$ \\\hline
$10^{-32}$ & $7482.7$ & $13867.7$ &  $0$ \\\hline
$10^{-31}$ & $7482.7$ & $13867.7$ &  $0$ \\\hline
$10^{-30}$ & $7482.7$ & $13867.7$ &  $0$ \\\hline
\end{tabular}\captionof{table}{\smaller Indicative values of the three form factor $b~(GeV^{-2})$, the  scalar field ($\varphi~(GeV)$), the GUT breaking vev of the scalar component of the superfield $\beta~(GeV)$ and the value of the scalar potential $V~(GeV^4)$ at the vacuum for the $G_2/SU(3)$ case.}\label{tab:case1}
\end{minipage}

\vspace{0.4cm}

From Figure \ref{fig:case1} we observe that for $b < 10^{-35}~GeV^{-2}$ the potential has a strongly negative minimum. This behaviour changes for $b > 10^{-35}~GeV^{-2}$, in which case a vanishing potential at the vacuum is obtained. Looking at Table \ref{tab:case1}, we also find a correlation between the critical point of $b$ and the sign of $\varphi$, which gets positive for the desired range of $b$.

\subsection{The $Sp_4/(SU(2)\times U(1))_{non-max}$ case}
In this case there are three scalar fields, namely the dilaton $\phi$, and the two radius moduli, $\varphi, \chi$. Nevertheless, this case is not genuinely nearly-K\"ahler, which means that we need to take the corresponding limit in which the two radii are equal, $R_1 = R_2$, which is translated to $\frac{\tilde{\varphi}}{\sqrt{2}} = \tilde{\chi}$. Taking this limit into consideration, the expressions of the fields, eq.\eqref{scalarscaseii} and that of the radius eq.\eqref{radiuscaseii} become:
\begin{align}
    \tilde{\phi} &= \tilde{\phi}(\phi, \varphi) = -\frac{1}{2}(\phi - \sqrt{3}\varphi)\,,\\
    \tilde{\varphi} &= \tilde{\varphi}(\phi, \varphi) = -\frac{1}{2}(\phi + \frac{1}{\sqrt{3}}\varphi)\,,
\end{align}
\begin{equation}
    R_1(\phi, \varphi) = re^{-\frac{\tilde{\varphi}(\phi, \varphi)}{2}}\,.
\end{equation} 
\\\\
\underline{\emph{The gravitational sector:}} \\\\
From eq.\eqref{gravitysectorcaseii}, the (negative-definite) contribution of the potential due to the gravitational sector is:
\begin{equation}
    V_{gr}(\phi, \varphi) = -\frac{15}{4\kappa^2}\frac{e^{-\tilde{\phi}(\phi,\varphi)}}{R_1(\phi,\varphi)^2}\,,
\end{equation}
which formally looks the same as in the previous case.
\\\\
\underline{\emph{The gauge sector:}} \\\\
The $R=(SU(2)\times U(1))_{non-max}$ is embedded into $G = E_8$ as:
\begin{align}
E_8 \supset & (SU(2)\times U(1))_{non-max} \times E_6\times U(1)\\
248 =& (3,1)_0 + (1,1)_0 + (1,78)_0 + (2,1)_3+ (2,1)_{-3}  \nonumber\\
&+ (1,27)_{-2} + (1,\bar{27})_{2} + (2,27)_1 + (2,\bar{27})_{-1}\nonumber
\end{align}
and according to the rule $G\supset H\times R$, the initial gauge group $E_8$ breaks to the subgroup $H=E_6\times U(1)$. The group $K$ in which the group $H=E_6\times U(1)$ breaks spontaneously is found in a more subtle way, given our choice that $S\subset G$. Therefore, we write down the decomposition of $G=E_8$ under one of its maximal subgroups:
\begin{align}
E_8&\supset SU(5)\times SU(5) \\
248 &= (1,24) + (24,1) + (5,\bar{10}) + (\bar{5},10) + (10,5) + (\bar{10},\bar{5})\,.
\end{align}
In turn, in order to find the group in which the subgroup $H = E_6\times U(1)$ breaks spontaneously we need to embed $S \subset G$. Given than $Sp_4$ is a maximal subgroup of $SU(5)$,
\begin{align}
    SU(5)&\supset Sp_4 \\ 
    24 &= 10 + 14\,,
\end{align}
we choose as $S$ the $Sp_4$ that is maximally embedded in $SU(5)$. Therefore the $Sp_4$ is embedded into the initial group, $G=E_8$ as $E_8\supset Sp_4 \times SU(5)$ and the final gauge group after spontaneous symmetry breaking is $K=SU(5)$. Now, in order that we understand the breaking pattern to $K$, we need to examine the decomposition of the 27-plet to the representations of the subgroups, which starts from the maximal decomposition:
\begin{align} 
E_6 &\supset SO(10) \times U(1) \nonumber\\
27 &= 1_4 + 10_{-2} + 16_1 \label{first_decomp}
\end{align}
and continues with:
\begin{align}
    SO(10) &\supset SU(5)\times U(1) \nonumber \\
    16 = 1_{-5} + \bar{5}_3 + &10_{-1}\,, \quad 10 = 5_2 + \bar{5}_{-2}\,.\label{second_decomp}
\end{align}
From the above, we consider that the components of the chiral superfields that break the symmetry of $H=E_6\times U(1)$ to $K=SU(5)$, i.e. the two singlet scalars, $\alpha, \beta$, acquire their vevs at the singlet direction of either the $16$, eq.\eqref{second_decomp}, or the $27$, \eqref{first_decomp} at the corresponding steps of the decomposition. Therefore, in the following calculations, we keep only the singlet directions, $\alpha, \beta$ and the contribution to the potential due to the gauge sector given in eq.\eqref{gaugesectorcaseii} becomes:
\begin{equation}
    V_{gauge}(\phi,\varphi,\alpha,\beta) = \frac{\alpha^\prime}{8 \kappa ^2}  e^{-\frac{1}{2}\tilde{\phi}(\phi,\varphi)} \left(-\frac{6 \alpha^2}{R_1(r,\phi ,\varphi )^2}-\frac{4 \beta ^2}{R_1(\phi ,\varphi )^2}+\frac{24}{R_1(\phi ,\varphi )^4}\right)\,.
\end{equation}
\\\\
\underline{\emph{The 3-form sector:}} \\\\
From eq.\eqref{3formsectorcaseii}, the contribution to the potential due to the 3-form part becomes:
\begin{equation}
    V_{3f}(\phi,\varphi,\alpha, \beta) = \frac{1}{4 \kappa ^2}e^{-\tilde{\phi}(\phi ,\varphi)}\left(\frac{\alpha ^{\prime 2}\left(\alpha^2+\beta ^2\right)^2}{R_1(\phi, \varphi )^2}+\frac{b^2}{R_1(\phi, \varphi )^6}\right)\,,
\end{equation}
where $\alpha, \beta$ are the components of the gauge fields that acquire vev after the spontaneous breaking of the subgroup $H=E_6\times U(1)$ to $K=SU(5)$, as explained above in the gauge sector and $b^2 = (2b^1 + b^2)^2$.

Putting the above three contributions together, the total scalar potential is found:
\begin{align}
    V &= \frac{e^{-\sqrt{3} \varphi -\phi }}{4 \kappa ^2 r^6} \left(b^2-r^2 e^{\frac{\varphi }{\sqrt{3}}+\frac{\phi }{4}} \left(\alpha^\prime  r^2 \left(3 \alpha^2+2 \beta ^2\right) e^{\frac{1}{4} \left(\sqrt{3} \varphi +2 \phi \right)}-\right.\right.\nonumber \\ &~~~~~~~~~~~~~~~~~~~~~~~~~~~ \left.\left.r^2 e^{\frac{3 \phi }{4}} \left(\alpha ^{\prime 2} a^4+2 \alpha ^{\prime 2} \alpha^2 \beta ^2+\alpha ^{\prime 2} \beta ^4-15\right)-12 \alpha  e^{\frac{\varphi}{4 \sqrt{3}}}\right)\right)\,.
\end{align}
Minimization of the above scalar potential is now done with respect to three fields, the $\varphi$ scalar field and the scalar components of the $\alpha, \beta$ fields that acquire vevs, again for  various values of $b$.\\

\begin{center}
\begin{minipage}[t]{0.7\textwidth}\centering
    \begin{tabular}{|l|r|r||r|r|r|}
\hline
$b$ & $\varphi$ & $V$ & $b$ & $\varphi$ & $V$ \\\hline
$10^{-38}$ & $-19.5$  & $-4.2 \cdot 10^{82}$ & $10^{4}$  & $268.5$  & $-9.2 \cdot 10^{-44}$ \\\hline
$10^{-36}$ & $-2.0$  & $-1.0 \cdot 10^{73}$ & $10^{7}$  & $268.5$  & $-7.4 \cdot 10^{-44}$ \\\hline
$10^{-31}$ & $268.5$  & $-9.2 \cdot 10^{-44}$ & $10^{8}$  & $273.0$  & $-3.1 \cdot 10^{-45}$ \\\hline
$10^{-17}$ & $268.5$  & $-9.2 \cdot 10^{-44}$& $10^{15}$  & $304.9$  & $-4.2 \cdot 10^{-55}$ \\\hline
$10^{-24}$ & $268.5$  & $-9.2 \cdot 10^{-44}$ & $10^{22}$  & $340.6$  & $-3.8 \cdot 10^{-66}$ \\\hline
$10^{-10}$ & $268.5$  & $-9.2 \cdot 10^{-44}$ & $10^{29}$  & $374.8$  & $-9.1 \cdot 10^{-77}$ \\\hline
$10^{-3}$  & $268.5$  & $-9.2 \cdot 10^{-44}$ & $10^{38}$  & $409.9$  & $-1.0 \cdot 10^{-87}$ \\\hline
\end{tabular}\captionof{table}{\smaller Indicative values of the three form factor $b~(GeV^{-2})$, the  scalar field ($\varphi~(GeV)$) and the value of the scalar potential $V~(GeV^4)$ at the vacuum for the $Sp_4/(SU(2)\times U(1))_{non-max}$ case.}\label{tab:case2}
\end{minipage}
\end{center}

From Table \ref{tab:case2} we are led to the conclusion that for all values of $b$, the minimum value of the potential is negative. As the orders of magnitude of $b$ increase, the value of the potential, from strongly negative approaches zero asymptotically, but never vanishes or gets positive.

\subsection{The $SU(3)/U(1)\times U(1)$ case}
This case is once again not genuinely nearly-K\"ahler, which means that we need to take the corresponding limit in which the three radii are equal, $R_1 = R_2 = R_3$, which effectively means $\tilde{\varphi} = \tilde{\chi} = \tilde{\psi}$. Taking this limit into consideration, the expressions for the scalars, eq.\eqref{scalarcaseiii} become:
\begin{align}
    \tilde{\phi} &= \tilde{\phi}(\phi, \varphi) = -\frac{1}{2}(\phi - \sqrt{3}\varphi)\,,\\
    \tilde{\varphi} &= \tilde{\varphi}(\phi, \varphi) = -\frac{1}{2}(\phi + \frac{1}{\sqrt{3}}\varphi)\,,
\end{align}
while the function of the radius with respect to the scalar fields, eq.\eqref{radiuscaseiii} is:
\begin{equation}
    R_1(\phi, \varphi) = re^{-\frac{\tilde{\varphi}(\phi, \varphi)}{2}}\,,
\end{equation}
where $r$ is a constant value as explained in the previous cases. 
\\\\
\underline{\emph{The gravitational sector:}} \\\\
The (negative-definite) contribution of the potential due to the gravitational part is:
\begin{equation}
    V_{gr}(\phi, \varphi) = -\frac{15}{4\kappa^2}\frac{e^{-\tilde{\phi}(\phi,\varphi)}}{R_1(\phi,\varphi)^2}\,,
\end{equation}
which formally looks the same as in the previous cases.
\\\\
\underline{\emph{The gauge sector:}} \\\\
The $R=U(1)\times U(1)$ is chosen to be embedded in $E_8$ as the maximal subgroup of $SU(3)$ in the decomposition of $E_8$ under its maximal subgroups,
\begin{align}
E_8&\supset SU(3)\times E_6~.\label{central3}
\end{align}
Then
\begin{align}
E_8 &\supset U(1)\times U(1) \times E_6\\
248 &= 1_{(0,0)} + 1_{(0,0)} + 1_{(3,\frac{1}{2})} + 1_{(-3,\frac{1}{2})} + 1_{(0,-1)} + 1_{(0,1)} + 1_{(-3,-\frac{1}{2})} + 1_{(3,-\frac{1}{2})} + 78_{(00)}  \nonumber\\
& + 27_{(3,\frac{1}{2})} + 27_{(-3,\frac{1}{2})} + 27_{(0,-1)}
+ \bar{27}_{(-3,-\frac{1}{2})} + \bar{27}_{(3,-\frac{1}{2})} + \bar{27}_{(0,1)}
\end{align}
and, according to the rule $G\supset H\times R$, the initial gauge group $E_8$ breaks to the subgroup $H=E_6\times U(1)\times U(1)$. Then the group $K$ in which the group $H$ breaks spontaneously is by construction (see eq.(\ref{central3})) the centralizer of $S$ in $G$, that is:
\begin{equation}
    K = C_{E_8}(SU(3)) = E_6\,.
\end{equation}
We consider that the breaking of $H=E_6\times U(1)\times U(1)$ to $K=E_6$ occurs after at least two out of the three scalar components of the chiral superfields that are singlets under the $E_6$ gauge group acquire vevs. Therefore, in the following calculations, we keep only the three singlet fields (considering the general case that all three get a vev) and, starting from eq.\eqref{gaugesectorcaseiii}, the contribution to the potential due to the gauge sector is:
\begin{align}
    V_{gauge}(\phi,\varphi,\alpha,\beta, \gamma) &= \frac{\alpha^\prime}{8 \kappa ^2}  e^{-\frac{\tilde{\phi}(\phi,\varphi)}{2}} \left(\frac{6}{5R_1(\phi ,\varphi)^4} -\frac{240 \sqrt{2} \left(a \beta  \gamma +a^* \beta ^* \gamma ^*\right)}{R_1(r,\phi ,\varphi )^2}\right.\nonumber\\
    &-\frac{4 \left(a a^*+\beta  \beta ^*+\gamma  \gamma ^*\right)}{R_1(r,\phi ,\varphi )^2}
    +40 \left(a \gamma  \alpha ^* \gamma ^*+a \beta  a^* \beta ^*+\beta  \gamma  \beta ^* \gamma ^*\right)^2\nonumber\\
    &\left.+\frac{5}{3} \left(a a^*+\beta  \beta ^*-2 \gamma  \gamma ^*\right)^2+15 \left(a a^*-\beta  \beta ^*\right)^2\right)\,.
\end{align}
\\\\
\underline{\emph{The 3-form sector:}} \\\\
From eq.\eqref{3formsectorcaseiii}, the contribution to the potential due to the 3-form sector becomes:
\begin{equation}
    V_{3f}(\phi,\varphi,\alpha, \beta, \gamma) = \frac{1}{4 \kappa ^2}e^{-\tilde{\phi}(\phi ,\varphi)}\left(\frac{\alpha ^{\prime 2}\left(\alpha^2+\beta ^2 + \gamma^2\right)^2}{R_1(\phi, \varphi )^2}+\frac{b^2}{R_1(\phi, \varphi )^6}\right)\,,
\end{equation}
where we have considered $b^2 = b_1^2 + b_2^2 + b_3^2$.

\noindent Putting the above three contributions together, the total scalar potential is found:
\begin{align}
V&=\frac{1}{8 \kappa ^2}e^{-\frac{2 \varphi }{\sqrt{3}}} \left(\frac{2 e^{-\frac{\varphi }{\sqrt{3}}-\phi } \left(\alpha ^2 r^4 e^{\frac{\varphi }{\sqrt{3}}+\phi } \left(a^2+\beta ^2+\gamma ^2\right)^2+b^2\right)}{r^6}\right.\nonumber\\
&\left.+\alpha  e^{\frac{5 \varphi }{4 \sqrt{3}}+\frac{\phi }{4}} \left(40 \left(\gamma  \gamma ^* \left(a \alpha ^*+\beta  \beta ^*\right)+a \beta  a^* \beta ^*\right)^2\right.\right. \nonumber\\
&\left.\left.+\frac{5}{3} \left(a a^*+\beta  \beta ^*-2 \gamma  \gamma ^*\right)^2+15 \left(a a^*-\beta  \beta ^*\right)^2-\frac{4 e^{\frac{1}{6} \left(-\sqrt{3} \varphi -3 \phi \right)} \left(a a^*+\beta  \beta ^*+\gamma  \gamma ^*\right)}{r^2}\right.\right. \nonumber\\
    &\left.\left.-\frac{240 \sqrt{2} e^{\frac{1}{6} \left(-\sqrt{3} \varphi -3 \phi \right)} \left(a \beta  \gamma +a^* \beta ^* \gamma ^*\right)}{r^2}+\frac{6 e^{-\frac{\varphi }{\sqrt{3}}-\phi }}{5 r^4}\right)-\frac{30}{r^2}\right)
\end{align}

This time the minimization of the scalar potential is done with respect to four fields, namely the $\varphi$ scalar field related to the radii and the  $\alpha, \beta, \gamma$ vev-acquiring scalar components of the respective superfields, once more for various values of the $b$ parameter.

The results found on Table \ref{tab:case3} lead us to the conclusion that for the various values of $b$ the behaviour is similar to the first case ($G_2/SU(3))$, but the strong change of the value of the potential  occurs at $b = 10^{-33}~GeV^{-2}$. For $b<10^{-33}~GeV^{-2}$ the behaviour of the potential at the vacuum is similar to the one demonstrated on the top diagram of Figure \ref{fig:case1}, with dependence on three gauge fields rather than one. For $b>10^{-33}~GeV^{-2}$ the bottom diagram of Figure \ref{fig:case1} describes the behaviour of the potential accurately enough.

\begin{center}
\begin{minipage}[t]{0.7\textwidth}\centering
    \begin{tabular}{|l|r|r|r|r|r|}
\hline
$b$ & $\varphi$ & <$\alpha$> & <$\beta$> & <$\gamma$> & $V$  \\\hline
$10^{-38}$ & $-19.6$ & $-12.6$ & $-6.8$ & $3.2$ & $-4.9 \cdot 10^{82}$ \\\hline
$10^{-37}$ & $-11.6$ & $-7.8$ & $-4.0$ & $2.1$ & $-4.8 \cdot 10^{78}$ \\\hline
$10^{-36}$ & $-3.6$ & $-2.2$ & $-1.0$ & $1.0$ & $-4.9 \cdot 10^{82}$ \\\hline
$10^{-35}$ & $4.5$ & $4.4$ & $1.4$ & $0.5$ & $-3.8 \cdot 10^{70}$ \\\hline
$10^{-34}$ & $12.8$ & $11.8$ & $2.2$ & $1.34$ & $-1.7 \cdot 10^{66}$ \\\hline
$10^{-33}$ & $2659.8$ & $2363.2$ & $159.8$ & $369.6$ & $0$ \\\hline
$10^{-32}$ & $2659.8$ & $2363.2$ & $159.8$ & $369.6$ & $0$ \\\hline
$10^{-31}$ & $2659.8$ & $2363.2$ & $159.8$ & $369.6$ & $0$ \\\hline
$10^{-30}$ & $2659.8$ & $2363.2$ & $159.8$ & $369.6$ & $0$ \\\hline
$10^{-29}$ & $2659.8$ & $2363.2$ & $159.8$ & $369.6$ & $0$ \\\hline
$10^{-28}$ & $2659.8$ & $2363.2$ & $159.8$ & $369.6$ & $0$ \\\hline
$10^{-27}$ & $2659.8$ & $2363.2$ & $159.8$ & $369.6$ & $0$ \\\hline
$10^{-26}$ & $2659.8$ & $2363.2$ & $159.8$ & $369.6$ & $0$ \\\hline
$10^{-25}$ & $2659.8$ & $2363.2$ & $159.8$ & $369.6$ & $0$ \\\hline
\end{tabular}\captionof{table}{\smaller Indicative values of the three form factor $b~(GeV^{-2})$, the  scalar field ($\varphi~(GeV)$), the GUT breaking vevs of the scalar components of the superfields $\alpha,~\beta$ and $\gamma~(GeV)$ and the value of the scalar potential $V~(GeV^4)$ at the vacuum for the $SU(3)/U(1)\times U(1)$ case.}\label{tab:case3}
\end{minipage}
\end{center}

\section{A few comments on the case of $SU(3)/U(1)\times U(1)$}

The dimensional reduction over the coset space $SU(3)/U(1)\times U(1)$ results in a 4-d gauge group that contains the promising $\mathcal{N}=1$ super trinification GUT as a subgroup. Thus, it is a strong candidate for the construction of phenomenologically viable models (see \cite{Manolakos:2020cco} and also \cite{Patellis:2020cue}). Therefore, it is important to point out that there are several differences regarding the parameters of the phenomenological models derived from this reduction and the present study, which serves more as a qualitative guide than a rigorous precision analysis.

In particular, a realistic approach to the above includes a further breaking by Wilson lines, in order to break our 4-d gauge group to the trinification group. That means that the manifold over which the original theory is dimensional reduced is upgraded to the multiply connected $SU(3)/U(1)\times U(1)\times Z_3$. Furthermore, in most versions of the phenomenological model, different fields acquire a vev at the unification scale. These changes may strongly affect the gauge sector of the scalar potential. 
As demonstrated in the previous section, the three-form sector is key, regarding the vanishing minimum of the total scalar potential, and the above changes are not expected to have a non-negligible impact on our calculation. One point that must be taken into consideration is the fact that in the present study we used the simplification that the string gauge coupling equals the unification gauge coupling, while a run of the renormalization group equations of the model between the compactification and the unification scale may alter the calculation above\footnote{Although a trinification model with three fermionic generations naturally features a vanishing 1-loop gauge $\beta$-function \cite{Ma:2004mi}, there are still corrections from higher loops.}. Additionally a different choice of scale for the radii can affect all three sectors of the potential. We are confident however, that all the above-mentioned changes will not have a qualitative impact on the calculation of the total scalar potential minimum, but rather a change of the $b$ condition of a few orders of magnitude. A last point is that the above minimization is done from a top to bottom point of view, and thus gives GUT breaking vevs at the order of 100-1000 GeV, rather than around $10^{16}$GeV. In a realistic approach, we observe that the demand that vevs are required at that high region relaxes the $b$ condition by several orders of magnitude.

\section{Conclusions}
In this work we examined the behaviour of the scalar potential of the 4-d theory that is obtained after the dimensional reduction of the 10-d $\mathcal{N}=1$ heterotic supergravity coupled to an $\mathcal{N}=1$ Yang-Mills sector over the three 6-d nearly-K\"ahler manifolds, namely the homogeneous 6-d non-symmetric coset spaces, $G_2/SU(3)$, $Sp(4)/SU(2)\times U(1)$ and $SU(3)/U(1)\times U(1)$.

According to our analysis we understand that in two out of three cases, $G_2/SU(3)$ and $SU(3)/U(1)\times U(1)$, there is a range of values of the parameter $b$ that accommodates a vanishing value of the minimum of the scalar potential, which means that vacua of Minkowski type are conditionally allowed. Also, for the second case, $Sp4/SU(2)\times U(1)$, for all values of $b$ the vacua remains of AdS type. None of the above cases produces vacua of dS type. 

All in all, the fact that we have obtained such a result for the $SU(3)/U(1)\times U(1)$ case is really important, taking into consideration that, in general, this is the case of solid phenomenological aspirations. The result for this case is more than welcome since it is the theory on which our 4-d phenomenological studies are based.

\section*{Acknowledgements}
We would like to thank Athanasios Chatzistavrakidis and Pantelis Manousselis for useful and constructive comments on the current work. Also, we are thankful to Dieter L\"ust for his support and constant encouragement.

GP has been supported by the Basic Research Programme, PEVE2020 of National Technical University of Athens, Greece. One of us (GZ) would like to thank the DFG Exzellenzcluster 2181:STRUCTURES of Heidelberg University, MPP-Munich, A.v.Humboldt Foundation and CERN-TH for support. Also, this work (GM) has been supported by the Polish National Science Center Grant No. 2017/27/B/ST2/02531.

\bibliographystyle{unsrt}

\bibliography{bibliography}
\end{document}